\documentclass[useAMS,usenatbib,referee]{biom}
\usepackage{amsmath,amssymb}
\newcommand{\abs}[1]{\left\vert#1\right\vert}

\newcommand{\what}[1]{\widehat{#1}}

\newcommand{\wtilde}[1]{\widetilde{#1}}

\newcommand{\qmq}[1]{\quad\mbox{#1}\quad}

\usepackage[figuresright]{rotating}

\title[Phase I Cancer Trial Designs]{Incorporating Individual and Collective Ethics into\\ Phase I Cancer Trial Designs}

\author{Jay Bartroff$^{*}$\email{bartroff@usc.edu}\\
           Department of Mathematics, University of Southern
           California,\\ 3620 South Vermont Avenue, KAP 108, Los
           Angeles, CA 90089, U.S.A.
            \and 
            Tze Leung Lai$^{**}$\\
            Department of Statistics, Sequoia Hall, Stanford
            University, Stanford, CA 94305, U.S.A.\\
$^{**}$\textit{email:} lait@stat.stanford.edu
           }

\begin{document}


\pagerange{\pageref{firstpage}--\pageref{lastpage}} 
\pubyear{2009}


\label{firstpage}


\begin{abstract} 
A general framework is proposed for Bayesian model-based designs of Phase~I cancer trials, in which a general criterion for coherence \citep{Cheung05} of a design is also developed.  This framework can incorporate both ``individual'' and ``collective'' ethics  into the design of the trial. We propose a new design which minimizes a risk function composed of two terms, with one representing the individual risk of the current dose and the other representing the collective risk. The performance of this design, which is measured in terms of the accuracy of the estimated target dose at the end of the trial, the toxicity and overdose rates, and certain loss functions reflecting the individual and collective ethics, is studied and compared with existing Bayesian model-based designs and is shown to have better performance than existing designs.
\end{abstract}

%
%


\maketitle


\section{Introduction}\label{sec:into} A Phase~I trial for a new treatment is generally intended to
determine a dose to use in subsequent
Phase~II and III testing. Phase~I cancer trials have the additional complexity that the treatment in question is usually a cytotoxic agent and the efficacy usually increases with dose, and therefore it is widely accepted that some degree of
toxicity must be tolerated to experience any substantial therapeutic
effects. Hence, an acceptable proportion $p$ of patients experiencing \textit{dose limiting toxicities} (DLTs) is generally agreed on before the trial, which depends on
the type and severity of the DLT; the dose resulting in this
proportion is thus referred to as the \textit{maximum tolerated dose} (MTD). In addition to the explicitly stated objective of determining the MTD, a Phase~I cancer trial also has the implicit goal of safe treatment of the patients in the trial. However, the aims of treating patients in the trial and generating an efficient design to estimate the MTD for future patients often run counter to each other. Commonly used designs in Phase~I cancer trials implicitly place their focus on the safety of the patients in the trial, beginning from a conservatively low starting dose and  escalating cautiously. Escalation is further slowed by the assignment of the same dose to groups of consecutive patients, as in the widely used 3-plus-3 design, which is convenient to administer and shortens trial duration by simultaneously following patients in groups of 3. \citet{VonHoff91}  have documented that the overall response rates in these Phase~I trials are low, and substantial numbers of patients are treated at doses that are retrospectively found to be non-therapeutic.  Moreover, as pointed out by \citet{OQuigley90}, these designs are very inefficient for estimating the MTD, which is implied by the 3-plus-3 design to correspond to the case $p=1/3$. They proposed a Bayesian model-based design, called the ``continual reassessment method'' (CRM), to choose the dose levels sequentially, making use of all past data at each stage.

More than ninety  new Phase~I methods were published between 1991 and 2006 \citep{Rogatko07}, and there have been several reviews of the new methods \citep[e.g.,][]{Rosenberger02}. In this paper we focus on Bayesian model-based designs and Section~\ref{sec:framework} describes a general framework to develop and analyze them. As shown in Sections~\ref{sec:framework} and \ref{sec:coh+1step}, this framework allows one to incorporate the competing aims of a Phase~I cancer trial by choosing the loss function accordingly. It also enables one to derive certain desirable properties of the design, such as coherence \citep{Cheung05}, from the loss function, or to enforce them by using simple reformulations in this framework. Section~\ref{sec:examples} provides implementation details and gives a simulation study comparing Bayesian designs that correspond to different loss functions in the setting of a colon cancer trial considered by \citet{Babb98}.

\section{Posterior Distributions, Loss Functions and Sequential Dose Determination}\label{sec:framework}

A commonly used model-based approach to Phase~I cancer clinical trial design assumes the usual logistic regression model for the probability~$F_\theta(x)$ of DLT at dose level~$x$:
\begin{equation}\label{eq:2PL}
F_\theta(x)=1/(1+e^{-(\alpha+\beta x)}),
\end{equation} in which $\beta>0$ and $\theta=(\alpha,\beta)$ is unknown and to be estimated from the observed pairs $(x_i,y_i)$, where $y_i=1$ if the $i$th subject, treated at dose~$x_i$, experiences DLT and $y_i=0$ otherwise. The frequentist approach to inference on  $\theta$ uses the likelihood function and estimates $\theta$ by maximum likelihood, while the Bayesian approach assumes a prior distribution of $\theta$ and uses the posterior distribution for inference on $\theta$.

Denote the MTD by $\eta=F_\theta^{-1}(p)$ and the posterior distribution of $\theta$ based on $(x_1,y_1),\ldots,(x_k,y_k)$ by $\Pi_k$, and let $\Pi_0$ denote the prior distribution. The Bayes estimate of $\eta$ with respect to squared error loss is the posterior mean $E_{\Pi_k}(\eta)$, and the CRM proposed by \citet{OQuigley90} uses this posterior mean to set the dose for the next patient, i.e., $x_{k+1}=E_{\Pi_k}(\eta)$.  Instead of the posterior mean, \citet{Babb98} proposed to set $x_{k+1}$ equal to the $\omega$-quantile of the posterior distribution, where $0<\omega<1/2$ is chosen to be slightly less than $p$ in their examples. This design is called ``escalation with overdose control'' (EWOC) and $\omega$ is called the ``feasibility bound.'' A sequence of doses $x_n$ is called ``Bayesian feasible'' at level $1-\omega$ if $P_{\Pi_{n-1}}(\eta\ge x_n)\ge 1-\omega$ for all $n\ge 1$, and the EWOC doses are optimal among Bayesian-feasible ones; see \citet{Zacks98}.

Note that the dose for the $n$th patient in CRM or EWOC depends only on the posterior distribution~$\Pi_{n-1}$, i.e., $x_n$ is a functional $f(\Pi_{n-1})$ of $\Pi_{n-1}$. This functional defines $\{\Pi_k: k\ge 0\}$ as a Markov chain whose states are distributions on the parameter space~$\Theta$ and whose state transitions are given by the following.

\bigskip

\noindent\textit{Bayesian updating scheme:} Given current state~$\Pi$ (which is a prior distribution of $\theta$), let $x=f(\Pi)$ and generate first $\theta$ from $\Pi$ and then $y\sim\mbox{Bern}(F_\theta(x))$. The new state is the posterior distribution of $\theta$ given $(x,y)$.

\bigskip

The functional $x=f(\Pi)$ for CRM is $E_\Pi(\eta)$, which minimizes the expected squared error loss $E_\Pi[(\eta-x)^2]$. As pointed out by 
\citet{Babb98}, the symmetric nature of the squared error loss may not be appropriate for modeling the toxic response to a cancer
treatment. Instead of squared error loss, EWOC with feasibility bound~$\omega$ uses the functional $x=x(\Pi)$ that minimizes the asymmetric loss function $E_\Pi[\ell(\eta,x)]$, where 
\begin{equation}\label{eq:EWOCloss}
\ell(\eta,x)=\left\{\begin{array}{ll} \omega(\eta-x),&\mbox{if $x\le
\eta$}\\ 
(1-\omega)(x-\eta),&\mbox{if $x\ge \eta$}.\end{array}\right.\end{equation}  More generally,  we can consider other loss functions $\ell(\theta,x)$ and define $x(\Pi)$ that attains $\min_xE_\Pi[\ell(\theta,x)]$. In particular, the following example gives a response-based version of EWOC.

\bigskip

\textit{Example 1: Inverted overdose control.}   The EWOC loss function (\ref{eq:EWOCloss}) penalizes an overdose $x>\eta$ by the amount $(1-\omega)(x-\eta)$, and an under-dose $x<\eta$ by the amount $\omega(x-\eta)$.  However, a dose~$x$ deemed ``too large'' on this scale may actually correspond to a probability of DLT not much larger than the target rate~$p$ depending on the dose-response curve, making $x$ a relatively desirable dose.  Likewise, a small value of $\abs{x-\eta}$ may correspond to a large discrepancy between the actual DLT probability~$F_\theta(x)$ and $p$. \citet{Hardwick01} suggest to measure  the excess/deficit of the DLT rate on the ``probability scale.'' Taking $0<\gamma<1/2$, this leads to the ``inverted'' loss function
\begin{equation}\label{eq:invEWOC}
\ell(\theta,x)=\left\{\begin{array}{ll} \gamma(p-F_\theta(x)),&\mbox{if $x\le 
\eta$}\\ 
(1-\gamma)(F_\theta(x)-p),&\mbox{if $x\ge \eta$}.\end{array}\right.\end{equation}

\bigskip

\noindent\textit{Remark:} The loss when the dose falls below $\eta$, measured by the difference $\omega(\eta-x)$ in (\ref{eq:EWOCloss}) and $\gamma(p-F_\theta(x))$ in (\ref{eq:invEWOC}), should ideally be measured by the difference in response rates at $\eta$ and $x$, respectively, when efficacy data, taking the value~$1$ if the patient responds to the treatment and $0$ otherwise,  are also available besides the toxicity data. Note, however, that this involves bivariate efficacy-toxicity data. While many existing designs solely consider toxicity outcomes and the MTD, designs that incorporate  efficacy responses as well have been proposed by a number of authors, including \citet{Li95}, \citet{Hardwick01}, \citet{Kpamegan01}, \citet{Thall04}, \citet{Dragalin06}, \citet{Dragalin08}, and \citet{Pronzato10}. When efficacy responses are available, the \textit{minimum effective dose} (MED) is of interest, i.e., the lowest dose at which some desired proportion of positive efficacy responses is attained.  When both efficacy and toxicity data are available, the \textit{optimal safe dose}, which is the dose between the MED and the MTD maximizing the probability of simultaneous efficacy and non-toxicity,  is of interest. Since this paper focuses on univariate toxicity data, we consider elsewhere better alternatives to (\ref{eq:invEWOC}) for $x\le \eta$ that also require efficacy data.

\bigskip

Noting that the explicitly stated objective of a Phase~I cancer trial is to estimate the MTD, \citet{Whitehead95} considered Bayesian sequential designs that are optimal, in some sense, for this estimation problem. \citet{Haines03} made use of the theory of optimal design of experiments \citep{Fedorov72,Atkinson92,Dette04} to construct Bayesian $c$- and $D$-optimal designs, and further imposed a relaxed Bayesian feasibility constraint on the design to avoid highly toxic doses. Optimal design theory involves a design measure~$\xi$ on the dose space~$\mathcal{X}$, and a sequential design updates the empirical design measure~$\xi_{n-1}$ at stage $n$ by changing it to $\xi_n$ with the addition of the dose~$x_n$. The empirical measure~$\xi_n$ of the doses $x_1,\ldots,x_n$ up to stage~$n$ can be represented by $\xi_n=n^{-1}\sum_{i=1}^n\delta_{x_i}$, where $\delta_x$ is the probability measure degenerate at $x$. We let $||\xi||$ denote the number of $x_i$ (not necessarily distinct) in the support of $\xi$. Thus $||\xi_n||=n$ and $||\xi_0||=0$, with $\xi_0$ being the zero measure on $\mathcal{X}$. To include the construction of sequential Bayesian optimal designs as a special case of our general approach, we can modify the preceding procedure that minimizes $E_\Pi[\ell(\theta,x)]$ to choose the next dose based on the current posterior distribution~$\Pi$, by including the current design measure~$\xi$ in the loss function.

\bigskip

\textit{Example 2: Bayesian $c$- or $D$-optimal designs.}  As described by \citet{Haines03}, optimal design theory is concerned with choosing a design measure~$\xi$ to minimize a convex function~$\Psi$ of the information matrix $M(\theta,\xi)=\int I(\theta,x)d\xi(x)$, where $I(\theta,x)$ is the Fisher information matrix at design point~$x$:
\begin{equation*}
I(\theta,x)=\frac{e^{\alpha+\beta x}}{(1+e^{\alpha+\beta x})^2}\left(\begin{array}{cc}
 1 &x  \\
x  & x^2
\end{array}\right).
\end{equation*} The convex function~$\Psi$ is associated with the optimality criterion, e.g., $\Psi(M)=-\log\det(M)$ for $D$-optimality and $\Psi(M)=c'M^{-1}c$ for $c$-optimality. Since $\theta=(\alpha,\beta)$ is unknown, the frequentist approach uses a sequential design that replaces $\theta$ in $M(\theta,\xi_t)$ by its maximum likelihood estimate at every stage~$t$. The Bayesian approach puts a prior distribution~$\Pi_0$ on $\theta$ and minimizes $\int \Psi(M(\theta,\xi_t))d\Pi_0(\theta)$. Noting that this Bayesian approach does not accomodate the fact that patients are assigned doses sequentially in Phase~I trials, \citet[][Section~5]{Haines03} propose to start the optimal design after an initial sample of $k$ patients so that the dose $x$ of a patient after this initial sample can be determined by minimizing
\begin{equation}\label{eq:seq_opt}
\int \Psi\left(\{k M(\theta,\xi_k)+I(\theta,x)\}/(k+1)\right) d\Pi_k(\theta),
\end{equation} where $\xi_k$ is the empirical measure of the initial sample of design points and $\Pi_k$ is the posterior distribution of $\theta$ based on the initial sample.

We can easily extend our loss function approach to Bayes sequential designs by including $\xi$ as an argument of the loss function in this setting. Let $\Pi$ be the current posterior distribution of $\theta$ and $\xi$ be the current empirical design  measure. Define 
\begin{equation}\label{eq:ell-opt}
\ell(\theta,x;\xi)=\Psi(M(\theta,\xi_{+\{x\}})),\qmq{where} \xi_{+\{x\}}=\frac{||\xi||\xi+\delta_x}{||\xi||+1}.
\end{equation} The sequential Bayes optimal design chooses the next design level $x$ that minimizes $E_\Pi\ell(\theta,x;\xi)$. The measure $\xi_{+\{x\}}$ in (\ref{eq:ell-opt}) represents the new empirical measure obtained by adding $x$ to the support of $\xi$, with $||\xi_{+\{x\}}||=||\xi||+1$. We can also impose a relaxed feasibility constraint in the choice of $x$: 
\begin{equation}\label{eq:bayes-opt}
\mbox{Minimize}\quad E_\Pi\ell(\theta,x;\xi)\qmq{subject to} P_\Pi(\wtilde{\eta}<x)\le\omega,
\end{equation} as in  \citet{Haines03}, where $\wtilde{\eta}=F_\theta^{-1}(q)$ with $q\ge p$ and $\omega$ is a prescribed positive constant. If $q=p$, then $\wtilde{\eta}=\eta$ and the constraint corresponds to requiring the doses to be Bayesian feasible (see the description of EWOC above).

\section{Coherence and Dilemma Between Individual and Collective Ethics}\label{sec:coh+1step}

The preceding section has focused on determining the next dose by minimizing  $E_\Pi[\ell(\theta,x)]$, where $\Pi$ is the current posterior distribution and $\ell$ is a loss function incorporating the trial's main objective into the Bayes sequential design. In Example~2 we have shown how additional information, such as the empirical measure of previous design points, can be included in the minimization problem to determine the dose. The following subsections extend this idea to address two important issues in Phase~I cancer clinical trial designs.

\subsection{Coherence and Its Enforcement}\label{sec:coherence}

Motivated by ethical concerns, \citet{Cheung05} introduced coherence principles for sequential dose escalation or de-escalation. A dose sequence  is said to be ``coherent'' if a higher  (respectively, lower) dose is not given to the next patient when the current patient experiences (respectively, does not experience) DLT. In particular, CRM and EWOC are coherent and the following theorem, whose proof is given in the Appendix, provides conditions for the coherence of a Bayes sequential design that minimizes the posterior loss at every stage.

\begin{theorem} \label{thm:coh} Suppose that the dose space is a finite interval and that $\ell(\eta,x)$ is convex in $x$ for every fixed $\eta$. Assume that for fixed $x>x'$, $\ell(\eta,x)-\ell(\eta,x')$ is non-increasing in $\eta$.  Then the dose sequence $x_n=\arg\min_x E_{\Pi_{n-1}}\ell(\eta,x)$ is coherent.
\end{theorem}

Theorem~\ref{thm:coh} shows that CRM is coherent since $\ell(\eta,x)=(\eta-x)^2$ is convex and $$\ell(\eta,x)-\ell(\eta,x')=-2\eta(x-x')+x^2-(x')^2$$ is non-increasing in $\eta$ for $x>x'$. The loss function~(\ref{eq:EWOCloss}) associated with EWOC also satisfies the assumption of Theorem~\ref{thm:coh}, which therefore shows the coherence of EWOC.  The loss functions in Examples~1 and 2, however, may not satisfy the assumptions of Theorem~\ref{thm:coh}. Moreover, a modification of EWOC recommended by its proponents \citep{Babb04}, in which the feasibility bound is escalated throughout the trial from a low starting value to $1/2$ at the end of the trial, does not satisfy the assumptions of Theorem~\ref{thm:coh} and it indeed exhibits slight incoherence in the simulation studies in Section~\ref{sec:examples}. This can be understood by noting that, toward the end of the trial, the posterior distribution does not change much from patient to patient, and that an increase in the feasibility bound may overwhelm the slight downward shift in the posterior following an outcome~$y=0$, causing a dose higher than the previous to be assigned. \citet[][p.~865]{Cheung05} also found a certain two-stage modification of CRM to be  incoherent.  On the other hand, we can enforce coherence by modifying $x_n=f(\Pi_{n-1})$ into $x_n=f(\Pi_{n-1},x_{n-1},y_{n-1})$, where
\begin{equation}
f(\Pi,x^*,y)=\begin{cases}
\arg\min_{x\le x^*} E_\Pi\ell(\eta,x)&\mbox{if $y=1$,}\\
\arg\min_{x\ge x^*} E_\Pi\ell(\eta,x)&\mbox{if $y=0$.}
\end{cases}
\end{equation}

\subsection{Treatment of Current Patient versus Information for Future Patients}\label{sec:1step}

We have noted in Section~\ref{sec:framework} that CRM or EWOC treats the next patient at the dose~$x$ that minimizes $E_\Pi[\ell(\theta,x)]$ for $\ell(\eta,x)$ given by $(\eta-x)^2$ or by (\ref{eq:EWOCloss}), where $\Pi$ is the current posterior distribution. This is tantamount to dosing the next patient at the best guess of $\eta$, where ``best'' means ``closest'' according to some measure of distance from $\eta$. On the other hand, a Bayesian $c$- or $D$-optimal design aims at generating doses that provide most information, as measured by the Fisher information matrix of a design measure, for estimating the dose-toxicity curve to benefit future patients. To resolve this dilemma between treatment of patients in the trial and efficient experimental design for post-trial parameter estimation, \citet{Bartroff10b} considered the finite-horizon optimization problem of choosing the dose levels $x_1,x_2,\ldots,x_n$ sequentially to minimize the ``global risk''
\begin{equation}\label{eq:fin_hor}
E_{\Pi_0}\left[\sum_{i=1}^n h(\eta,x_i)+g(\what{\eta}_n,\eta)\right],
\end{equation} in which $\Pi_0$ denotes the prior distribution of $\theta$, $h(\eta,x_i)$ represents the loss for the $i$th patient in the trial, $\what{\eta}_n$ is the terminal estimate of the MTD and $g$ represents a terminal loss function. The optimizing doses~$x_i$  depend on $n-i$, where the horizon~$n$ is the sample size of the trial, and therefore are not of the form $x_i=f(\Pi_{i-1})$ considered in Section~\ref{sec:framework}. In terms of ``individual'' and ``collective'' ethics, note that (\ref{eq:fin_hor}) measures the individual effect of the dose $x_k$ on the $k$th patient through $h(\eta,x_k)$, and its collective effect on future patients through $\sum_{i>k} h(\eta,x_i)+g(\what{\eta}_n,\eta)$.

By using a discounted infinite-horizon version of (\ref{eq:fin_hor}), we can still have solutions of the form $x_i=f(\Pi_{i-1})$ for some functional~$f$ that only depends on $\Pi_{i-1}$. Specifically, take a discount factor $0<\delta<1$ and replace (\ref{eq:fin_hor}) by
\begin{equation}\label{eq:globrisk}
E_{\Pi_0}\left[\sum_{i=1}^\infty h(\eta,x_i)\delta^{i-1}\right]
\end{equation}  as the definition of global risk. Note that this global risk measures the individual effect of the dose $x_k$ on the $k$th patient through $h(\eta,x_k)$, and its collective effect on future patients through $\sum_{i>k} h(\eta,x_i)\delta^{i-k}$. This means the myopic dose~$x_k$ that minimizes $E_{\Pi_{k-1}}[h(\eta,x)]$ for treating the $k$th patient has to be perturbed such that it also helps to create a more informative posterior distribution $\Pi_k$ that is used for dosing future patients. Note that (\ref{eq:globrisk}) does not have the term $g(\what{\eta}_n,\eta)$ appearing in the finite-horizon problem~(\ref{eq:fin_hor}), but even without this term, the global risk~(\ref{eq:globrisk}) still captures the collective effect of the doses, as indicated above. As we have pointed out in Section~\ref{sec:framework}, if $x_i$ is of the form $f(\Pi_{i-1})$ for all $i$, then $\{\Pi_k: k\ge 0\}$ is a Markov chain whose states are distributions of $\theta$ and undergo Markovian dynamics described by the updating scheme for posterior distributions.  In the context of the present problem of minimizing (\ref{eq:globrisk}), the optimal expected loss~$V(\Pi)$ at state~$\Pi$ (posterior distribution of $\theta$) satisfies Bellman's dynamic programming equation
\begin{equation}\label{eq:Bellman}
V(\Pi)=\inf_x E_\Pi\{h(\eta,x)+\delta E_\Pi V(\Pi_{+\{x\}})\},
\end{equation} where $\Pi_{+\{x\}}$ is the new posterior distribution of $\theta$ after $(x,y)$ is observed, with $y\sim\mbox{Bern}(F_\theta(x))$ and $\theta\sim\Pi$; see the Bayesian updating scheme in Section~\ref{sec:framework}. For finite-state controlled Markov chains, iteration is a commonly used method to solve (\ref{eq:Bellman}); see \citet[][Section~1.3]{Bertsekas07}. In the present case, not only is the state space infinite, but it is also infinite-dimensional (space of all posterior distributions of $\theta$), making dynamic programming intractable. 

The main complexity of the infinite-horizon problem is that the dose~$x$ for the next patient involves also consideration for future patients who will receive optimal doses themselves; these future doses depend on the future posterior distributions. A simple way to reduce the complexity is to consider two (instead of infinitely many) future patients. This amounts to choosing the next dose~$x$ to minimize $E_\Pi\ell(\eta,x;\Pi)$ when the current posterior distribution of $\theta$ is $\Pi$, where
\begin{equation}\label{eq:1-stp-ell}
\ell(\eta,x;\Pi)=h(\eta,x)+\lambda E_\Pi\{E_\Pi[h(\eta',x')|x_1=x, y_1]\},
\end{equation} in which $\eta'=F_{\theta'}^{-1}(p)$ with $\theta'\sim\Pi'$, and $\Pi'$ and $x'$ are defined below. The first summand in (\ref{eq:1-stp-ell}) measures the (toxicity) effect of the dose~$x$ on the patient receiving it. The second summand considers the patient who follows and receives a myopic dose~$x'$ which minimizes the patient's posterior loss; the myopic dose is optimal because there are no more patients involved in (\ref{eq:1-stp-ell}). The effect of $x$ on this second patient is through the posterior distribution~$\Pi'$ that updates $\Pi$ after observing $(x_1,y_1)$, with $x_1=x$. Since $y_1$ is not yet observed, the expectation outside the curly brackets is taken over $y_1\sim\mbox{Bern}(F_\theta(x))$, with $\theta\sim\Pi$. For example, when implemented with $h(\eta,x)$ given by the EWOC loss function~(\ref{eq:EWOCloss}), this proposal can be viewed as a modification of EWOC since it utilizes its loss function but adds an additional term to represent the effect on future patients.
 
 Unlike $0<\delta<1$ in the discounted infinite-horizon problem, the choice of $\lambda>0$ in (\ref{eq:1-stp-ell}) can exceed 1 and reflects the balance between the collective ethics in generating information for future patients and the individual ethics for the patient receiving the dose. Although we use here a single patient to represent all patients following the one receiving the next dose, because the posterior distributions also change successively, the doses are functionals of these posterior distributions. 
 
\section{Implementation and a Simulation Study}\label{sec:examples}
In this section we first describe three main components in the implementation of the above Bayesian sequential designs and then evaluate their performance in a simulation study.

\subsection{Updating the Posterior Distribution}

Letting $\eta$ denote the MTD and $\rho=F_\theta(x_{\min})$, we follow \citet{Babb98} to transform $(\alpha,\beta)$ in~(\ref{eq:2PL}) to~$(\rho,\eta)$ via the formulas
\begin{eqnarray*}
\alpha&=&\frac{x_{\min}\log(p^{-1}-1)-\eta\log(\rho^{-1}-1)}{\eta-x_{\min}},\\
\beta&=&\frac{\log(\rho^{-1}-1)-\log(p^{-1}-1)}{\eta-x_{\min}},
\end{eqnarray*}
and therefore
$$\alpha+\beta x = \frac{(x-\eta)\log(\rho^{-1}-1)-(x-x_{\min})\log(p^{-1}-1)}{\eta-x_{\min}}=G(x,\rho,\eta).$$
We assume that the joint prior distribution of $(\rho,\eta)$ has density $\pi(\rho,\eta)$ with support on $[0,p]\times [x_{\min},x_{\max}]$. Therefore the $\mathcal{F}_{k-1}$-posterior density of $(\rho,\eta)$ is
\begin{equation}\label{eq:post}
\pi_{k-1}(\rho,\eta)=C\prod_{i=1}^{k-1}\left[\frac{1}{1+e^{-G(x_i,\rho,\eta)}}\right]^{y_i} \left[\frac{1}{1+e^{G(x_i,\rho,\eta)}}\right]^{1-y_i} \pi(\rho,\eta),\end{equation}
  where $$C^{-1}= \int_{x_{\min}}^{x_{\max}} \int_0^q \prod_{i=1}^{k-1}\left[\frac{1}{1+e^{-G(x_i,\rho,\eta)}}\right]^{y_i} \left[\frac{1}{1+e^{G(x_i,\rho,\eta)}}\right]^{1-y_i} \pi(\rho,\eta) d\rho d\eta.$$ The marginal $\mathcal{F}_{k-1}$-posterior distribution of $\eta$ is then $\int_0^p \pi_{k-1}(\rho,\eta) d\rho$, and the CRM and EWOC doses based on $\mathcal{F}_{k-1}$ are the mean and $\omega$-quantile of this distribution, respectively.  

\subsection{Computation of $E_\Pi\ell(\eta,x)$ and its Minimizer in Sections~\ref{sec:framework} and \ref{sec:coherence}}


The integrals in (\ref{eq:post}) can be evaluated by using a numerical double-integration routine involving Gaussian quadrature in \textsf{MATLAB}. This can be used to evaluate $E_\Pi\ell(\eta,x)$ for a posterior distribution~$\Pi$. We can find the minimum of $E_\Pi\ell(\eta,x)$ over $x$ by a grid search in $[x_{\min},x_{\max}]$, or by using gradient descent if $\ell$ is smooth. For computation of the constrained Bayesian optimal design (\ref{eq:bayes-opt}), a constrained nonlinear optimization routine in \textsf{MATLAB} can be used in conjunction with numerical integration, as outlined in \citet[][p.~593]{Haines03}.
 
\subsection{Minimization of $E_\Pi\ell(\eta,x;\Pi)$ in Section~\ref{sec:1step}}

While MCMC or rejection sampling can be used to compute (\ref{eq:1-stp-ell}) for any candidate dose~$x$, importance sampling \citep[e.g.,][Chapter~3.3]{Robert04} is a simple, robust alternative that takes advantage of the fact that just an expectation with respect to the posterior distribution is needed. Letting $\Pi_0$ denote the uniform distribution of the transformed coordinates~$(\rho,\eta)$ over $[0,p]\times [x_{\min},x_{\max}]$, we have 
\begin{equation}\label{eq:impsamp1}
E_\Pi\ell(\eta,x;\Pi)\approx B^{-1}\sum_{b=1}^B\ell(\eta_b,x;\Pi)\frac{\pi(\rho_b,\eta_b)}{\pi_0(\rho_b,\eta_b)}
\end{equation} for large $B$, where $(\rho_b,\eta_b)$, $b=1,\ldots,B$,  are i.i.d.\ and generated from $\Pi_0$. Letting $\Pi_{+\{x,y\}}$ denote the posterior distribution obtained from $\Pi$ by including $(x,y)$ and letting $x'=x'(\Pi_{+\{x,y\}})$, the nested expectation in (\ref{eq:1-stp-ell}) can be similarly approximated by using
\begin{gather}\label{eq:impsamp2}
E_\Pi[h(\eta',x')|x_1=x,y]=E_{\Pi_{+\{x,y\}}}h(\eta',x')\approx B^{-1}\sum_{b=1}^B h(\eta_b',x') \frac{\pi_{+\{x,y\}}(\rho_b',\eta_b')}{\pi_0(\rho_b',\eta_b')},\\
\label{eq:impsamp3}
P_\Pi(y=1|x)=\int F_\theta(x)d\Pi(\theta)\approx  B^{-1}\sum_{b=1}^B F_{\theta_b''}(x) \frac{\pi(\rho_b'',\eta_b'')}{\pi_0(\rho_b'',\eta_b'')},
\end{gather} where $(\rho_b',\eta_b')$, $(\rho_b'',\eta_b'')\sim\Pi_0$ and $\theta_b''=\theta(\rho_b'',\eta_b'')$. Let $H_\Pi(x,y)$ and $Q_\Pi(x)$ denote the right-hand sides of (\ref{eq:impsamp2}) and (\ref{eq:impsamp3}), respectively. Combining (\ref{eq:impsamp1})-(\ref{eq:impsamp3}) gives
\begin{equation}\label{eq:1stprsk}
E_\Pi\ell(\eta,x;\Pi)\approx B^{-1}\sum_{b=1}^B\left\{h(\eta_b,x)+\lambda\left[H_\Pi(x,0) (1-Q_\Pi(x))+ H_\Pi(x,1) Q_\Pi(x)\right]\right\}\frac{\pi(\rho_b,\eta_b)}{\pi_0(\rho_b,\eta_b)}.
\end{equation} We can minimize the right-hand side of (\ref{eq:1stprsk}) over $x\in[x_{\min},x_{\max}]$ by using a bounded minimization routine in \textsf{MATLAB}.

\subsection{Simulation Study}\label{sec:sim}

To compare the proposed procedure in Section~\ref{sec:1step} to EWOC, CRM, and the inverted overdose control  (IVOC) design in Example~1, a simulation study was performed in the setting of the trial to determine the MTD of the antimetabolite 5-fluorouracil (5-FU) for treating solid tumors in the colon, as described in \citet{Babb98}. Based on previous studies of 5-FU, a dose of 140 mg/m$^2$ of 5-FU was believed to be safe, and the MTD was believed to be no greater than 425 mg/m$^2$, thus the dose space was taken to be the interval $[x_{\min},x_{\max}]=[140,425]$. The two-parameter logistic model~(\ref{eq:2PL}) was chosen based on previous experience with the agent, and the uniform distribution over $[0,p]\times [x_{\min}, x_{\max}]$ was chosen as the prior distribution~$\Pi_0$ for $(\rho,\eta)$, with $p=1/3$. The feasibility bound of $\omega=.25$ was chosen, which was also used here for the IVOC weight~$\gamma$ in (\ref{eq:invEWOC}).  In a trial of length $n=24$, Table~\ref{table:sim} compares EWOC that uses a linearly escalated feasibility bound \citep{Babb04}, denoted by EWOC$^*$, with IVOC, CRM, and the proposed design in Section~\ref{sec:1step} with $h$ in (\ref{eq:1-stp-ell}) given by the EWOC loss function (and denoted by EWOC$_+$, in which $+$ signifies an additional future patient considered by (\ref{eq:1-stp-ell})), for two different values of the discount factor~$\lambda$ in (\ref{eq:1-stp-ell}). Each entry in the table was calculated from 10,000 simulated trials. The first set of rows is a Bayesian setting in which, for each replication, a pair $(\rho,\eta)$ is drawn from $\Pi_0$, and the next three sets of rows are frequentist settings (denoted Freq$_1$, Freq$_2$, Freq$_3$) where the true values $(\rho,\eta)$ are set at fixed values for all 10,000 replications; these three pairs of fixed values were drawn from $\Pi_0$.  A comprehensive comparison of EWOC, CRM, sequential $c$-optimal, constrained $D$-optimal, ADP and other  designs has been given by \citet{Bartroff10b}, who use approximate dynamic programming (ADP) to minimize the finite-horizon risk~(\ref{eq:fin_hor}).

\begin{table}\caption{Risk$_1$, Risk$_2$, bias and RMSE of the final MTD estimate, DLT rate, MTD overdoes rate (OD), excess DLT rate $E[F_\theta(x)-p]^+$ (OD$^*$), and coherence violation rate (ChV), with SEs in parentheses, of various designs.}\label{table:sim}
\begin{tabular}{l|lllll}\hline\hline
Statistic&EWOC$^*$&IVOC&CRM&EWOC$_{+,\lambda=.1}$ &EWOC$_{+,\lambda=.4}$ \\\hline \hline
\multicolumn{6}{c}{Bayesian: $(\rho,\eta)\sim\Pi_0$}\\\hline
Risk$_1$&485.5 (3.6)&723.2 (3.6)&986.1 (45.9)&469.5 (3.0)&454.8 (2.8) \\
Risk$_2$&1.03 (.01)&1.44 (.02)&1.54 (.02)&.85 (.01)&.73 (.007)\\
Bias&-9.67 (.6)&-50.2 (.8)&22.3 (1.6)&2.42 (.7)&-5.9 (.6)\\
RMSE&61.5 (.7)&75.8 (.2)&157.3 (.8)&66.9 (.2)&58.6 (.7)\\
DLT (\%)&33.5 (.001)&26.6 (9$\times 10^{-4}$ )&39.1 (.001)&29.3 (.0009)& 29.1 (9$\times 10^{-4}$)\\
OD (\%)&37.4 (.001)&17.5 (8$\times 10^{-4}$ )&55.6 (.001)&29.6 (9$\times 10^{-4}$)&27.0 (9$\times 10^{-4}$)\\
OD$^*$&.043 (2$\times 10^{-4}$)&.043 (3$\times 10^{-4}$)&.078 (3$\times 10^{-4}$)&.029 (2$\times 10^{-4}$)&.021 (1$\times 10^{-4}$)\\
ChV (\%)&0 (0)&.6 (.04)&0 (0)&15.0 (8$\times 10^{-4}$)&14.4 (8$\times 10^{-4}$)\\
\hline
\multicolumn{6}{c}{Freq$_1$: $\rho=.07, \eta=403.9$}\\\hline
Risk$_1$&585.6 (1.3)&1302.2 (.8)&368.8 (1.1)&175.9 (1.0)&170.3 (1.1)\\
Risk$_2$&.78 (.002)&1.40 (6$\times 10^{-4}$)&.53 (.001)&.29 (.001)&.25 (.001)\\
Bias&-51.1 (.2)&-151.7 (.2)&-30.1 (.4)&-23.2 (.3)&-48.7 (.2)\\
RMSE&58.8 (.8)&156.1 (.6)&44.9 (.1)&49.0 (.5)&19.2 (.4)\\
DLT (\%)&20.6  (8$\times 10^{-4}$)&10.3 (7$\times 10^{-4}$)&24.9 (9$\times 10^{-4}$)&18.7 (8$\times 10^{-4}$)&18.6 (8$\times 10^{-4}$)\\
OD (\%)&0 (0)&0 (0)&2.3 (3$\times 10^{-4}$)&1.2 (2$\times 10^{-4}$)&1.1 (2$\times 10^{-4}$)\\
OD$^*$&0 (0)&0 (0)&4$\times 10^{-4}$ (1$\times 10^{-5}$)&.001 (1$\times 10^{-5}$)&.001 (1$\times 10^{-5}$)\\
ChV (\%)&0 (0)&0 (0)&0 (0)&0 (0)&0 (0)\\
\hline
\multicolumn{6}{c}{Freq$_2$: $\rho=.19, \eta=269.1$}\\\hline
Risk$_1$&402.8 (.8)&423.1 (.9)&313.7&275.0 (3.7)&266.2 (3.8)\\
Risk$_2$&.53 (.003)&.49 (.001)&.99 (.006)&.43 (.001)&.26 (.001)\\
Bias&15.0 (.5)&-28.9 (.1)&32.5 (.4)&17.3 (.7)& 12.3 (.4)\\
RMSE&47.5 (.2)&32.0 (.7)&52.1 (.7)&38.2 (.3)&25.2 (.3)\\
DLT (\%)&32.6 (.001)&25.4 (9$\times 10^{-4}$)&38.2 (.001)&28.2 (9$\times 10^{-4}$)&24.8 (9$\times 10^{-4}$)\\
OD (\%)&32.8 (.001)&0 (0)&83.1 (8$\times 10^{-4}$)&27.1 (9$\times 10^{-4}$)&26.9 (9$\times 10^{-4}$)\\
OD$^*$&.02 (7$\times 10^{-5}$)&0 (0)&.053 (8$\times 10^{-5}$)&.03 (4$\times 10^{-5}$)&.01 (4$\times 10^{-5}$)\\
ChV (\%)&4.3 (4$\times 10^{-4}$)&3.1 (4$\times 10^{-4}$)&0 (0)&17.6 (8$\times 10^{-4}$)& 12.7 (7$\times 10^{-4}$)\\ 
\hline
\multicolumn{6}{c}{Freq$_3$: $\rho=.30, \eta=226.7$}\\\hline
Risk$_1$&674.3 (4.9)&232.1 (2.9)&1444.9 (6.3)&158.9 (4.9)& 146.1 (3.5)\\
Risk$_2$&.27 (.002)&.09 (4$\times 10^{-4}$)&.59 (.003)&.05 (4$\times 10^{-4}$)&.05 (6$\times 10^{-4}$)\\
Bias&42.9 (.62)&10.3 (.14)&81.4 (.5)&54.9 (.54)& 47.4 (.49)\\
RMSE&74.2 (.7)&17.4 (.5)&96.8 (.8)&18.0 (.5)&13.2 (.2)\\
DLT (\%)&34.2 (.001)&32.7 (.001)&36.6 (.001)&35.8 (.001)&35.7 (.001)\\
OD (\%)&73.2 (.001)&54.3 (.001)&93.5 (5$\times 10^{-4}$)&88.8 (6$\times 10^{-4}$)&81.0 (6$\times 10^{-4}$)\\
OD$^*$&.014 (3$\times 10^{-5}$)&.002 (2$\times 10^{-5}$)&.032 (4$\times 10^{-5}$)&.031 (4$\times 10^{-5}$)&.030 (4$\times 10^{-5}$)\\
ChV (\%)&0 (0)&3.9 (5$\times 10^{-4}$)&0 (0)&10.0 (6$\times 10^{-4}$)&9.6 (6$\times 10^{-4}$)\\
\hline\hline
\end{tabular}
\end{table}

Table~\ref{table:sim} reports two different risk measures. Since the length of the trial is fixed at $n=24$ in the simulation study, Risk$_1$ is the finite-horizon analog of (\ref{eq:globrisk}), which is (\ref{eq:fin_hor}) with $h$ given by the EWOC loss function (\ref{eq:EWOCloss}) and no terminal loss (i.e., $g=0$), and Risk$_2$ is the same risk function but with $h$ given by the ``inverted'' loss function~(\ref{eq:invEWOC}). Also reported are the bias and RMSE (root mean squared error $\{E(\what{\eta}_n-\eta)^2\}^{1/2}$) of the terminal MTD estimate~$\what{\eta}_n$ (which is the mean of the terminal posterior distribution of $\eta$), the DLT rate~$P(y=1)$ (denoted DLT), the overdose rate $P(x>\eta)$ (denoted OD), the excess DLT rate $E[F_\theta(x)-p]^+$ (denoted OD$^*$), and the coherence violation rate (denoted ChV) $$(n-1)^{-1}\sum_{i=1}^{n-1} \{P(y_i=0, x_{i+1}<x_i)+P(y_i=1, x_{i+1}>x_i)\}.$$ In these expressions, $P$ and $E$ denote the probability and expectation, respectively, with respect to the prior distribution in the Bayesian setting, or with respect to the appropriate fixed values of $(\rho,\eta)$ in the frequentist settings, and are computed by Monte Carlo.

In terms of Risk$_1$ and Risk$_2$, EWOC$_+$ performs better in the Bayesian setting  than the myopic designs EWOC$^*$, IVOC, and CRM, in that order.  This occurs in the frequentist settings as well, although the ordering of the myopic designs varies depending on the particular parameter values. Even though it myopically minimizes the posterior risk at every stage, IVOC performs poorly in terms of the cumulative risk, Risk$_2$, in the Bayesian setting.  A possible explanation is that its loss function (\ref{eq:invEWOC}) is a function of $F_\theta(x)$, whose posterior distribution (induced by the posterior distribution of $\theta$) has relatively large variance toward the middle of the interval $(0,1)$ in which $F_\theta(x)$ takes values and, in particular, near $p=1/3$, resulting in low initial doses observed in the simulations. On the other hand, in the Freq$_3$ setting, where $\eta$ is relatively small and the dose-response curve is relatively flat (e.g., $\rho$ large), IVOC performs well in terms of the risks. In terms of estimation,  EWOC$_{+,\lambda=.4}$ has the smallest RMSE, with EWOC$^*$ and EWOC$_{+,\lambda=.1}$ both comparable in the Bayesian setting. Moreover, EWOC$_{+,\lambda=.4}$ has uniformly the smallest RMSE in the frequentist settings, with IVOC comparable to it in Freq$_2$ and IVOC and EWOC$_{+,\lambda=.1}$ comparable to it in Freq$_3$.

\section{Conclusion and Discussion}\label{sec:conc}

In this paper we present a general formulation of Bayesian sequential design of Phase~I cancer trials.  This formulation enables us to prove a general coherence result in Theorem~\ref{thm:coh} applicable to any design that can be defined as the minimizer of the posterior risk when the loss function satisfies some mild conditions.  Although the theorem is proved for the widely-used logistic regression model~(\ref{eq:2PL}), the last paragraph of its proof in the Appendix shows that it is applicable to any dose-response model that is non-increasing in the MTD, such as the model $F_\theta(x)=\{(\tanh x+1)/2\}^\theta$, which is also popular.

In Section~\ref{sec:1step}  we propose a new design that incorporates both the individual ethics of the current patient begin administered the dose, through a given loss function such as the EWOC loss~(\ref{eq:EWOCloss}), and the collective ethics of all future patients by including an additional term in the overall loss function to represent the dose's information content  for determining another dose for the next patient.  The simulation study in Section~\ref{sec:sim} shows that this new design is indeed an improvement over myopic designs in terms of global risk minimization, post-trial estimation of the MTD, and DLT and OD rates. This design provides a practical alternative to the optimal design associated with the intractable Markov decision problem of minimizing (\ref{eq:globrisk}), which requires at each stage the daunting consideration of all future posterior distributions and calculating their associated optimal doses. For the finite-horizon problem of minimizing (\ref{eq:fin_hor}), \citet{Bartroff10b} have developed an approximate solution which is a time-varying mixture of myopic and $c$-optimal designs. The new design in Section~\ref{sec:1step}, which can be described by a time-invariant functional of the posterior distribution at each stage,  is substantially simpler computationally and provides substantial improvement over the myopic designs. We conjecture that with suitably chosen $\lambda$ (depending on $\delta$), its global risk~(\ref{eq:globrisk}) can approximate that of the optimal design minimizing (\ref{eq:globrisk}). Instead of minimizing (\ref{eq:globrisk}) directly, it may be possible to obtain a good lower bound for (\ref{eq:globrisk}). Such a bound, which can provide a benchmark for assessing the proposed design, is a topic for future work.

We also consider an ``inverted'' loss function~(\ref{eq:invEWOC}), which measures deviation from the target DLT rate~$p$ on the probability scale rather than on the dose scale, and the associated myopic design IVOC.  Even though IVOC minimizes the myopic posterior expected loss~(\ref{eq:invEWOC}) at each stage, its cumulative global loss Risk$_2$ in Table~\ref{table:sim} is far from optimum, exceeding even that of EWOC which uses a completely different loss function, in the Bayesian setting. On the other hand, the design proposed in Section~\ref{sec:1step} can be applied with the IVOC loss function~(\ref{eq:invEWOC}) to yield a substantially improved design IVOC$_+$. 

\backmatter

\section*{Appendix}

\textit{Proof of Theorem~\ref{thm:coh}.} We prove coherence in de-escalation; the proof for escalation is similar. Let $x_{\min}<x_{\max}$ be the boundaries of the dose space, which is assumed to be a finite interval. For fixed $\eta$, since $\ell(\eta,x)$  is a convex function of $x$, its right derivative~$\ell_x(\eta,x)$ with respect to $x$ is nondecreasing for $x_{\min}\le x<x_{\max}$, and the same is also true for the left derivative for $x_{\min}< x\le x_{\max}$. Moreover, the left and right derivatives are equal and continuous except for at most countably many points; see \citet[][pages~214, 228, 244]{Rockafellar70}. Let $x_\Pi=\arg\min_x E_\Pi\ell(\eta,x)$, $\wtilde{\Pi}$ be the posterior distribution obtained from $\Pi$ and the additional dose-response pair $(x,y)=(x_\Pi,1)$, and let $L(x)=E_{\wtilde{\Pi}}\ell(\eta,x)$. Since $\ell(\eta,x)$ is convex in $x$ for every $\eta$, so is $L(x)$; moreover, its right derivative is given by $\dot{L}_+(x)= E_{\wtilde{\Pi}}\ell_x(\eta,x)$. To show that $x_{\wtilde{\Pi}}\le x_\Pi$, we shall assume that $x_\Pi<x_{\max}$ because the case $x_\Pi=x_{\max}$ is trivial. It suffices to show that $\dot{L}_+(x_\Pi)\ge 0$ because $L$ is convex and has minimizer $x_{\wtilde{\Pi}}$. Since $E_\Pi\ell_x(\eta,x_\Pi)\ge 0$ and $d\wtilde{\Pi}(\theta)=F_\theta(x_\Pi)d\Pi(\theta)/\int F_{\theta'}(x_\Pi)d\Pi(\theta')$, recalling that $(x,y)=(x_\Pi,1)$, it follows that
\begin{align}
\dot{L}_+(x_\Pi)&\ge E_{\wtilde{\Pi}}\ell_x(\eta,x_\Pi)- E_{\Pi}\ell_x(\eta,x_\Pi)\nonumber\\
&=\frac{\int\ell_x(\eta,x_\Pi)F_\theta(x_\Pi)d\Pi(\theta)}{\int F_{\theta'}(x_\Pi)d\Pi(\theta')}-\frac{\int\ell_x(\eta,x_\Pi) d\Pi(\theta)}{\int d\Pi(\theta')}\nonumber\\
&=A\left/\int F_{\theta'}(x_\Pi)d\Pi(\theta')\right.,\label{eq:Ldot}
\end{align} where $A=\int\int \ell_x(\eta,x_\Pi)[F_{\theta}(x_\Pi)-F_{\theta'}(x_\Pi)]d\Pi(\theta) d\Pi(\theta')$. A change of variables also yields $A=-\int\int \ell_x(\eta',x_\Pi)[F_{\theta}(x_\Pi)-F_{\theta'}(x_\Pi)]d\Pi(\theta) d\Pi(\theta')$. Hence
\begin{equation}\label{eq:2A}
2A=\int\int [\ell_x(\eta,x_\Pi)-\ell_x(\eta',x_\Pi)][F_{\theta}(x_\Pi)-F_{\theta'}(x_\Pi)]d\Pi(\theta) d\Pi(\theta')\ge 0,
\end{equation} in which the inequality follows from
\begin{equation}\label{eq:int-pos}
[\ell_x(\eta,x_\Pi)-\ell_x(\eta',x_\Pi)][F_{\theta}(x_\Pi)-F_{\theta'}(x_\Pi)]\ge 0
\end{equation} for all $x$, $\theta$ and $\theta'$, as will be shown below. Combining (\ref{eq:Ldot}) and (\ref{eq:2A}) yields $\dot{L}_+(x_\Pi)\ge 0$, completing the proof of the theorem. 

From the assumption that $\ell(\eta,x)-\ell(\eta,x')$ is non-increasing in $\eta$ for any $x>x'$, it follows that $\ell_x(\eta,x)$ is non-increasing in $\eta$ for fixed $x$. It therefore suffices for the proof of (\ref{eq:int-pos}) to show that $F_\theta(x)$ is non-increasing in $\eta=F_\theta^{-1}(p)$. Since $p^{-1}=1/F_\theta(\eta)=1+e^{-(\alpha+\beta\eta)}$, $F_\theta(x)=1/[1+\exp\{\log(p^{-1}-1)+\beta\eta-\beta x\}]$, which is non-increasing in $\eta$ since $\beta>0$.


\section*{Acknowledgments}
This work was supported in part by National
Science Foundation grants DMS-0907241  at University of Southern California and DMS-0805879 at Stanford University.  The authors thank the Associate Editor and two referees for their helpful comments.





\begin{thebibliography}{}

\bibitem[\protect\citeauthoryear{Atkinson and Donev}{1992}]{Atkinson92}
Atkinson, A.~C. and Donev, A.~N. (1992).
\newblock {\em Optimum Experimental Designs}.
\newblock Oxford University Press.

\bibitem[\protect\citeauthoryear{Babb and Rogatko}{2004}]{Babb04}
Babb, J. and Rogatko, A. (2004).
\newblock Bayesian methods for cancer phase {I} clinical trials.
\newblock In Geller, N.~L., editor, {\em Advances in Clinical Trial
  Biostatistics}. Marcel Dekker.

\bibitem[\protect\citeauthoryear{Babb, Rogatko and Zacks}{1998}]{Babb98}
Babb, J., Rogatko, A., and Zacks, S. (1998).
\newblock Cancer phase {I} clinical trials: Efficient dose escalation with
  overdose control.
\newblock {\em Statistics in Medicine} {\bf 17,} 1103--1120.

\bibitem[\protect\citeauthoryear{Bartroff and Lai}{2010}]{Bartroff10b}
Bartroff, J. and Lai, T.~L. (2010).
\newblock Approximate dynamic programming and its applications to the design of
  phase {I} cancer trials.
\newblock {\em Statistical Science}, in press.

\bibitem[\protect\citeauthoryear{Bertsekas}{2007}]{Bertsekas07}
Bertsekas, D.~P. (2007).
\newblock {\em Dynamic Programming and Optimal Control}, volume~2.
\newblock Athena Scientific, Belmont, MA, 3rd edition.

\bibitem[\protect\citeauthoryear{Cheung}{2005}]{Cheung05}
Cheung, Y.~K. (2005).
\newblock Coherence principles in dose-finding studies.
\newblock {\em Biometrika} {\bf 92,} 863--873.

\bibitem[\protect\citeauthoryear{Dette, Melas and Pepelyshev}{2004}]{Dette04}
Dette, H., Melas, V.~B., and Pepelyshev, A. (2004).
\newblock Optimal designs for a class of nonlinear regression models.
\newblock {\em The Annals of Statistics} {\bf 32,} 2142--2167.

\bibitem[\protect\citeauthoryear{Dragalin and Fedorov}{2006}]{Dragalin06}
Dragalin, V. and Fedorov, V. (2006).
\newblock Adaptive designs for dose-finding based on efficacy-toxicity
  response.
\newblock {\em Journal of Statistical Planning and Inference} {\bf 136,}
  1800--1823.

\bibitem[\protect\citeauthoryear{Dragalin, Fedorov and Wu}{2008}]{Dragalin08}
Dragalin, V., Fedorov, V., and Wu, Y. (2008).
\newblock Adaptive designs for selecting drug combinations based on
  efficacy-toxicity response.
\newblock {\em Journal of Statistical Planning and Inference} {\bf 138,}
  352--373.

\bibitem[\protect\citeauthoryear{Fedorov}{1972}]{Fedorov72}
Fedorov, V.~V. (1972).
\newblock {\em Theory of Optimal Experiments}.
\newblock Academic Press, New York.

\bibitem[\protect\citeauthoryear{Haines, Perevozskaya and
  Rosenberger}{2003}]{Haines03}
Haines, L.~M., Perevozskaya, I., and Rosenberger, W.~F. (2003).
\newblock {B}ayesian optimal design for phase {I} clinical trials.
\newblock {\em Biometrics} {\bf 59,} 591--600.

\bibitem[\protect\citeauthoryear{Hardwick and Stout}{2001}]{Hardwick01}
Hardwick, J. and Stout, Q.~F. (2001).
\newblock Optimizing a unimodal response function for binary variables.
\newblock In Atkinson, A., Bogacka, B., and Zhigljavsky, A., editors, {\em
  Optimum Design 2000}, pages 195--210. Kluwer Academic Publishers, Dordrecht.

\bibitem[\protect\citeauthoryear{Kpamegan and Flournoy}{2001}]{Kpamegan01}
Kpamegan, E.~E. and Flournoy, N. (2001).
\newblock An optimizing up-and-down design.
\newblock In Atkinson, A., Bogacka, B., and Zhigljavsky, A., editors, {\em
  Optimum Design 2000}. Kluwer Academic Publishers, Dordrecht.

\bibitem[\protect\citeauthoryear{Li, Durham and Flournoy}{1995}]{Li95}
Li, Z., Durham, S.~D., and Flournoy, N. (1995).
\newblock An adaptive design for maximization of a contingent binary response.
\newblock In Flournoy, N. and Rosenberger, W.~F., editors, {\em Adaptive
  Designs}, pages 179--196. Institute of Mathematical Statistics.

\bibitem[\protect\citeauthoryear{O'Quigley, Pepe and Fisher}{1990}]{OQuigley90}
O'Quigley, J., Pepe, M., and Fisher, L. (1990).
\newblock Continual reassessment method: {A} practical design for phase {I}
  clinical trials in cancer.
\newblock {\em Biometrics} {\bf 46,} 33--48.

\bibitem[\protect\citeauthoryear{Pronzato}{2010}]{Pronzato10}
Pronzato, L. (2010).
\newblock Penalized optimal designs for dose-finding.
\newblock {\em Journal of Statistical Planning and Inference} {\bf 140,}
  283--296.

\bibitem[\protect\citeauthoryear{Robert and Casella}{2004}]{Robert04}
Robert, C.~P. and Casella, G. (2004).
\newblock {\em {M}onte {C}arlo Statistical Methods}.
\newblock Springer-Verlag, New York, 2nd edition.

\bibitem[\protect\citeauthoryear{Rockafellar}{1970}]{Rockafellar70}
Rockafellar, R.~T. (1970).
\newblock {\em Convex Analysis}.
\newblock Princeton Mathematical Series, No. 28. Princeton University Press,
  Princeton, N.J.

\bibitem[\protect\citeauthoryear{Rogatko, Schoeneck, Jonas, Tighiouart, Khuri
  and Porter}{2007}]{Rogatko07}
Rogatko, A., Schoeneck, D., Jonas, W., Tighiouart, M., Khuri, F., and Porter,
  A. (2007).
\newblock Translation of innovative designs into phase {I} trials.
\newblock {\em Journal of Clinical Oncology} {\bf 25,} 4982--4986.

\bibitem[\protect\citeauthoryear{Rosenberger and Haines}{2002}]{Rosenberger02}
Rosenberger, W.~F. and Haines, L.~M. (2002).
\newblock Competing designs for phase {I} clinical trials: A review.
\newblock {\em Statistics in Medicine} {\bf 21,} 2757--2770.

\bibitem[\protect\citeauthoryear{Thall and Cook}{2004}]{Thall04}
Thall, P.~F. and Cook, J.~D. (2004).
\newblock Dose-finding based on efficacy-toxicity trade-offs.
\newblock {\em Biometrics} {\bf 60,} 684--693.

\bibitem[\protect\citeauthoryear{Von~Hoff and Turner}{1991}]{VonHoff91}
Von~Hoff, D. and Turner, J. (1991).
\newblock Response rates, duration of response, and dose response effects in
  phase {I} studies of antineoplastics.
\newblock {\em Investigational New Drugs} {\bf 9,} 115--122.

\bibitem[\protect\citeauthoryear{Whitehead and Brunier}{1995}]{Whitehead95}
Whitehead, J. and Brunier, H. (1995).
\newblock Bayesian decision procedures for dose determining experiments.
\newblock {\em Statistics in Medicine} {\bf 14,} 885--893.

\bibitem[\protect\citeauthoryear{Zacks, Rogatko and Babb}{1998}]{Zacks98}
Zacks, S., Rogatko, A., and Babb, J. (1998).
\newblock Optimal {B}ayesian-feasible dose escalation for cancer phase {I}
  trials.
\newblock {\em Statistics \& Probability Letters} {\bf 38,} 215--220.

\end{thebibliography}

\def\cprime{$'$}

 
 \label{lastpage}


\end{document}